\newcommand{\beq}{\begin{equation}}
\newcommand{\eeq}{\end{equation}}
\newcommand{\beqns}{\begin{equation}}
\newcommand{\eeqns}{\end{equation}}
\newcommand{\beqar}{\begin{eqnarray}}
\newcommand{\bs}{\begin{eqnarray*}}
\newcommand{\eeqar}{\end{eqnarray}}
\newcommand{\es}{\end{eqnarray*}}
\newcommand{\beqml}{\begin{mathletters}}
\newcommand{\eeqml}{\end{mathletters}}
\newcommand{\x}{\mbox{\boldmath $x$}}
\newcommand{\0}{\mbox{\boldmath $x$}_0}
\newcommand{\y}{\mbox{\boldmath $y$}}

\newcommand{\Gtilde}{\tilde{G}}

\newcommand{\Ptilde}{\tilde{P}}

\newcommand{\intpi}{\int_{-\pi}^{+\pi} \frac{dq}{2\pi}}

\newfont{\fancy}{msbm10 scaled\magstep1}

\documentstyle[aps,preprint,eqsecnum]{revtex}
\begin{document}
\draft
\preprint{PURD-TH-93-07}

%
%

\title{ A new finite-size scaling approach to random walks}
\author{Achille Giacometti \footnote{present address: Istituto
di Fisica "Galileo Galilei", via Marzolo 8, 35131 Padova, Italy}
and Hisao Nakanishi}
\address{ Department of Physics, Purdue University,
West Lafayette, IN 47907 }
\date{\today}
\maketitle
\begin{abstract}
We present a new finite-size scaling method for the random walks (RW)
superseeding a previously widely used renormalization group approach,
which is shown here to be inconsistent.
The method is valid in any dimension and is based on the exact
solution for the two-point correlation function and on finite size scaling.
As an application, the phase diagram is derived for random walks with a
surface-bulk interaction where the system has either a surface or a defect.
Possible extensions to disordered systems are also discussed.
\end{abstract}
\vspace{5mm}
%
%
%
%
%
\newpage
\narrowtext
\section{Introduction}
\label{sec:intro}
Renormalization Group (RG) has been a cornerstone in the analytical
evaluation of the critical exponents of various statistical models
in the past two decades \cite{Wilson,Creswick,Zinn-Justin}.
Its application to self-avoiding walks (SAW) was a natural consequence
of the formulation of SAW as $n \rightarrow 0$ limit of the $n$-vector
model whose two-point correlation function yields, in this limit, the
generating function of the random walk with the excluded volume
effect \cite{Burkhardt}.

An extension of this approach to the case without self-avoidance was
carried out by Family and Gould (FG) \cite{FG} in the absence of
disorder, and by Sahimi and Jerauld (SJ) \cite{SJ} and
by Gould and Kohin (GK) \cite{GK}
in the disordered case, where the disorder was mimicked by
a site (or bond) percolation cluster.
The results in the latter cases were found to be in very good
agreement with those of numerical simulations \cite{HK}
of random walks on a percolating cluster on
the square lattice. However unlike GK who used the {\em kinetic} rule
for the random walk (more precisely they solved the model which
is nicknamed {\em myopic ant}) \cite{deG}, SJ used the {\em static}
recipe which is now recognized to correspond to the so-called
{\em ideal chain} model \cite{Maritan,GNMF,GM,GMN,GNF}.
Recent analysis \cite{Maritan,GNMF,GM,GMN,GNF} with various different
approaches have shown that the ideal chain does not belong
to the same universality class as the kinetic walks;
rather it is equivalent to the random walk in a trapping environment
\cite{Maritan,GMN}.  In view of this, we felt that a better
understanding of the method in the absence of disorder was needed.

By calculating exactly the two-point correlation functions
for a random walk on a quadrant of the lattice (the so-called
{\em corner rule} \cite{Burkhardt}), we will show that the
procedure introduced by FG needs to be revised because of the
uncontrolled approximations. This is also supported by
an exact analytical calculation on the one-dimensional analog.
We will then proceed to show how one can obtain a consistent
procedure based on the finite size scaling hypothesis \cite{Binder}.

The outline of this paper is as follows: In Sec.\ref{sec:two-point}
we first recall the derivation of the formal solution of the two-point
correlation function for the random walk. This solution constitutes
the basis of the real-space renormalization group approach
to this problem. The standard procedure is then reviewed and its
fundamental problems are discussed.  In Sec.\ref{sec:one-dimensional},
the one-dimensional version of this problem is
analyzed exactly in detail, and the inconsistencies of previous
renormalization approaches are exposed.
In Sec.\ref{sec:finite_size}, we introduce our new method
which is based entirely on the widely accepted finite size scaling
approach.  Then in Sec.\ref{sec:bulk-surface} this method is applied
to describe the phase diagram for a problem with different fugacities
for the {\it surface} and {\it bulk}.  The results are consistent with
mean field theory where applicable. Finally Sec.\ref{sec:conclusions}
contains some conclusions and perspectives on the extension of this
method to disordered systems.

\section{Two-point correlation function for random walks}
\label{sec:two-point}
It is well known that the unconstrained random walk on a lattice can
be solved by using a generating function technique pioneered by
Montroll \cite{MW-65,Barber-Ninham,ID}.  Let $P_{\0,\x}(N)$
the probability for the walker to be at the position
$\x \in $  { \fancy Z}$^d$ at the (discrete) time $N$, given
that it started at the site $\0$ at the initial time $0$.
The master equation to be solved is then
\beqar
P_{\0,\x}(N+1)&=&\frac{1}{z} \sum_{\y(\x)}P_{\0,\y}(N)
\label{ME_hyp}
\eeqar
where the notation $\y(\x)$ means that the sum is restricted to
the nearest neighbors of $\x$.
The number $C_{\0,\x}(N)=z^N P_{\0,\x}(N)$ of $N$-step walks
with end points $\0$ and $\x$ then satisfies the analogous equation,
\beqar
C_{\0,\x}(N+1)&=&\sum_{\y(\x)}C_{\0,\y}(N) \;\;.
\label{ME_chain}
\eeqar
In order to solve the master equation, it proves convenient
to introduce the generating function $G_{\0,\x}(k)$,
\beqar \label{Generating}
G_{\0,\x}(k)&=&\sum_{N=0}^{\infty} k^N C_{\0,\x}(N) \nonumber \\
            &=& \sum_{w:\0 \rightarrow \x} k^{|w|}
\eeqar
where $w$ is a walk having $\0$ and $\x$ as the end points, and $|w|$
is the number of steps associated with it.

By multiplying (\ref{ME_chain}) by $k^{N+1}$ and summing over all $N$,
one gets, taking into account the initial condition
$C_{\0,\x}(0) =\delta_{\0,\x}$,
\beqar \label{G_equation}
G_{\0,\x}(k)= k \sum_{\y(\x)} G_{\0,\y}(k) + \delta_{\0,\x} \;\;.
\eeqar
It is easy to see that $G_{\0,\x}(k)$ is also the two-point
correlation function of a scalar free-field theory
and that Eq.\ (\ref{Generating}) can be recovered from a
von Neumann expansion (see e.g. \cite{GMN,ID}).

Generally the procedure for a RG includes two basic steps:
First one coarse grains microscopic details in real space
or integrates over the {\em fast} modes in momentum space.
This is followed by a rescaling of the space/momentum and of the
model variables while conserving the partition function and
recasting the Hamiltonian in the same functional form as before.

In the case of the random walk, the requirement of the conservation of
the partition function amounts to a mapping between the
rescaled and original fugacities which can be formally expressed as:
\beqar \label{weights_map}
P(\{k'\},w') &=& \sum_{w | w'} P( \{k\},w)
\eeqar
where $k'$ and $w'$ are the fugacity and walk on the rescaled lattice,
respectively, $P$ stands for the partition function, and the sum is
constrained to all $w$ consistent with $w'$.  In the case
of self-avoiding walks, this procedure leads to a well known
polynomial recursion relation between $k'$ and $k$ whose linearization
around the fixed point leads to the value of the correlation length
exponent $\nu$. On the other hand, once the self-avoidance is
turned off, the polynomial recursion becomes an infinite series since
there are an infinite number of walks even in the smallest possible
cell. One is thus faced with the problem of finding either a way
of summing over an infinite number of walks or a truncation procedure.
Some time ago Family and Gould \cite{FG} devised a recipe along
the latter line. Their idea was that if $L=ba$ ($a$ being the lattice
constant) is the size of the system, then walks with
number of steps $N$ larger than $N_{MAX}$ given by
\beqar \label{r2_max_1}
N_{MAX} \sim [<R_N^2>]_{MAX} \sim L^2
\eeqar
will give a negligible contribution to the sum in
Eq.\ (\ref{weights_map}).

An $L \times L$ cell can be mapped into an $L' \times L'$ cell
by the requirement (\ref{weights_map}) as
\beqar \label{rel_chi}
{\chi}_{\hat{n}}(k',\frac{1}{L'}) &=& \chi_{\hat{n}}(k,\frac{1}{L})
\eeqar
where we defined the quantity
\beqar \label{chi_definition}
\chi_{\hat{n}}(k,\frac{1}{L}) &\equiv& \sum_{N=0}^{\infty}
C_{\hat{n}}(\0,N) k^N
\eeqar
and $C_{\hat{n}}(\0,N)$ is the number of $N$-step walks in the cell
starting from $\0$ and exiting in the direction $\hat{n} \;\;(n=1,2)$.
According to the assumption of FG, $\chi$ can be approximated as
\beqar \label{chi_FG}
\chi_{\hat{n}}(k,\frac{1}{L}) &\approx& \sum_{N=N_{MIN}}^{N_{MAX}}
C_{\hat{n}}(\0,N) k^N
\eeqar
where $N_{MAX}$ is related to the system size by Eq.\ (\ref{r2_max_1})
and $N_{MIN}$ is the minimum number of steps needed to reach the
{\it closest} exit of the $L \times L$ cell.

As an example (cf. Fig.\ref{fig1}), if we take $L'=a$ and $L=2a$, then
we have from Eqs.\ (\ref{rel_chi}) and (\ref{chi_FG}):
\beqar \label{rg_map_app}
k' &\approx& \sum_{N=N_{MIN}}^{N_{MAX}} C_{\hat{n}}(\0,N) k^N
=k^2+2k^3+5k^4+14k^5  \;\;.
\eeqar

On the other hand the total number of $N$-step walks having $\0,\x$
as the end points can be calculated easily from Eq.\ (\ref{G_equation})
which gives the quantity (\ref{Generating}) exactly.
Therefore both sides of Eq. (\ref{rel_chi}) can be calculated exactly
without any truncation procedure.
In Fig.\ref{fig2} we compare the right-hand side of
Eq.\ (\ref{chi_definition}) calculated exactly with the one calculated
using the FG truncation procedure,
which can clearly be recovered upon numerical Taylor expansion
of the exact result up to the desired order.
It is apparent that although in general the FG truncation procedure
seems to reproduce rather well the trend of the fixed point $k^*$,
it fails to reproduce the singularity present in (\ref{chi_definition}).
This singularity moves closer and closer
to the fixed point as the cell size increases.
The physical origin of this singularity stems from the fact that,
unlike other systems where criticality is reached only in the
infinite volume limit, the random walk has a criticality
in any finite cell, by taking the limit $N \rightarrow \infty$.
This has a consequence, as seen in Table \ref{table1},
that the value of the exponent $\nu$ overshoots the exact value
$1/2$ already at a very small cell size. This would also
be the case with the FG truncation recipe if the size of
the cell were pushed to a sufficiently large value (although the
behavior of such an approximation scheme for very large cell size
is not known and may be complicated, see \cite{HN}).

In the next section we will see how the same trend is found
in the one-dimensional analog, where everything can be calculated
analytically for any value of the system size $L$.

\section{The one-dimensional problem}
\label{sec:one-dimensional}
In order to gain firm understanding of the problems associated with
the standard renormalization approach, in this section we solve
a one-dimensional version of the aforementioned {\it corner rule}
renormalization method.  Subsequently a semi-infinite one-dimensional
model in the presence of an infinite, repulsive barrier
at the origin will also be discussed.

\subsection{Transfer matrix solution for the one-dimensional corner rule}
\label{sec:1d_corner}
Let us consider a one-dimensional lattice where $x=0,1,2,...,L$ and
the sites $x=-1$, $L+1$ have an infinitely repulsive barrier
(see Fig.\ \ref{fig3}).  (The lattice constant $a$ is set equal to $1$
for simplicity.)  The analog of Eq.\ (\ref{G_equation}) for the
correlation function for $x \geq 1$ is
\beqar \label{G_bulk}
G_{0,x}(k) &=& k [G_{0,x-1}(k)+G_{0,x+1}(k)]
\eeqar
along with the boundary conditions:
\beqar \label{bc1}
G_{0,0} (k) &=& k G_{0,1} (k) +1
\eeqar
and
\beqar \label{bc2}
G_{0,L} (k) &=& k G_{0,L-1}(k) \;\;,
\eeqar
where we have assumed that all walks start from $x=0$.

We can put this equation in a transfer matrix form,
\beqar \label{transfer_eq}
\Psi_x(k) &=& {\bf T} \Psi_{x+1}(k) \;\;,\;\; x=1,2,...,L-1
\eeqar
where we defined
\beqar \label{matrix_def}
{\bf T} &=&
\: \left(
\begin{array}{cc}
0 & 1 \\
-1 & \frac{1}{k}
\end{array} \;\; \right)\;,
\; \Psi_x = \left(
\begin{array}{c}
G_{0,x}(k) \\
G_{0,x-1}(k)
\end{array} \;\; \right)
\eeqar
The eigenvalues of the matrix ${\bf T}$ are clearl
\beqar \label{eigenvalues_T}
\lambda_{\pm} &=& \frac{1 \pm \sqrt{1-4k^2}}{2k}
\eeqar
which are real if $0 < k \leq 1/2$ and form a complex conjugate pair
if $1/2 < k <1$
(we consider $k <1$ in order to make the generating functions sensible).
Note that $\lambda_{+} \cdot \lambda_{-} =1$ in both cases.
The right $\{ { U }_{\alpha}\}$ and left $\{ { U}^{\alpha} \}$
eigenvectors ($\alpha= \pm$), are given by:
\beqar \label{eigenvectors_right_left}
{ U}_{\pm} &=&
\: \left(
\begin{array}{c}
1 \\
\lambda_{\pm}
\end{array} \;\; \right)\;,
\; {U}^{\pm} = \pm \frac{1}{\lambda_{+}-\lambda_{-}} \left(
\begin{array}{c}
- \lambda_{\mp} \\
1
\end{array} \;\; \right)
\eeqar
where we normalized so that $< U^{\alpha} | U_{\beta} > =
\delta^{\alpha}_{\beta}$.

Then expanding $\Psi_L(k)$ in terms
of the right eigenvectors and using the boundary condition (\ref{bc2})
we get the coefficients of $U_{\pm}$,
\beqar \label{coefficients_1d}
c_{L+1}^{\pm} &=& \pm \frac{G_{0,L}}{\lambda_{+}- \lambda_{-}} \;\;.
\eeqar
Expressing $\Psi_1(k)$ by iteration of Eq.\ (\ref{transfer_eq}) and
using (\ref{coefficients_1d}), we get
\beqar \label{G0_G1}
G_{0,1} &=& \frac{G_{0,L}}{\lambda_{+}- \lambda_{-}}
(\lambda_{+}^{L} - \lambda_{-}^{L}) \;\;, \\
G_{0,0} &=& \frac{G_{0,L}}{\lambda_{+}- \lambda_{-}}
(\lambda_{+}^{L+1} - \lambda_{-}^{L+1})  \;\;,
\eeqar
which determine $G_{0,L}$ if we further impose the other
boundary condition (\ref{bc1}). Thus we find,
\beqar \label{GL}
G_{0,L}(k) &=& \frac{\lambda_{+}-\lambda_{-}}{\lambda_{+}^{L+1}-
\lambda_{-}^{L+1}- k(\lambda_{+}^{L}-\lambda_{-}^{L})}
\eeqar
for the system size $L=1,2,...$.

Note that, although this solution is valid for arbitrarily large $L$,
the boundary condition (\ref{bc2}) makes the system {\em finite}.
This distinction will become more clear in the next subsection.

Once $G_{0,L}(k)$ has been obtained, the one-dimensional analog of
Eq.\ (\ref{chi_definition}),
namely $\chi_{\hat{1}}(k,1/L) \equiv G_{0,x}(k)|_{x=L}$, gives
the recursion relation corresponding to (\ref{rel_chi}):
\beqar \label{rel_G_G'_2}
\chi_{\hat{1}}(k',1/L') &=& \chi_{\hat{1}}(k,1/L)
\eeqar
The fixed point $k^*(L,L')$ is obtained by setting $k=k'=k^*$ in
(\ref{rel_G_G'_2}).
Using (\ref{GL}) and (\ref{rel_G_G'_2}) for renormalization from a cell
of size $L+1$ to that of size $L$, we get an implicit solution for
the fixed point $k^{*}(L+1,L)$,
\beqar \label{fixed_point_general}
k_L^* \equiv k^*(L+1,L) &=& \frac{\lambda_{+}^{* L+2}-\lambda_{-}^{* L+2}
- ( \lambda_{+}^{* L+1} -\lambda_{-}^{* L+1})}
{\lambda_{+}^{* L+1}-\lambda_{-}^{* L+1} - ( \lambda_{+}^{* L}
-\lambda_{-}^{* L})}
\eeqar
where $\lambda_{\pm}^{*}$ is $\lambda_{\pm}$ evaluated at $k=k_L^{*}$.

We expect $k_L^*$ to approach $1/2$ as $L \rightarrow \infty$ because
the critical fugacity $k_c$ for the infinite, one-dimensional problem
is exactly $1/2$.  Given this limit, we now look for the $L$ dependence
of $k_L^*$ for large but finite $L$. For this purpose,
we need to distinguish
the following two cases due to the square root in the expression
(\ref{eigenvalues_T}) for $\lambda$:

\begin{flushleft}
(a) Case $0 < k_L^* \leq 1/2$.
\end{flushleft}

If this were the case, the eigenvalues $\lambda_{\pm}^*(k)$ would be
real. By introducing an auxiliary variable $\alpha$ by
$\tanh \alpha = \sqrt{1-4 k_L^{*2}}$ and expressing
(\ref{fixed_point_general}) in terms of hyperbolic functions, we get
\beqar \label{fixed_point_a}
1/2 = \frac{(1+\tanh \alpha )^{L+2} - (1-\tanh \alpha )^{L+2}
      - [ (1+\tanh \alpha )^{L+1}-(1-\tanh \alpha )^{L+1} ]\cosh \alpha}
           {(1+\tanh \alpha )^{L+1} - (1-\tanh \alpha )^{L+1}
      - [ (1+\tanh \alpha )^{L}-(1-\tanh \alpha )^{L} ]\cosh \alpha} \;.
\eeqar
We see that the right hand side of (\ref{fixed_point_a}) tends to $1$
as $\alpha \rightarrow 0$ (i.e., $k_L^*\rightarrow (\frac{1}{2})^{-}$).
This shows that $k_L^*$ cannot be in this range at least for large $L$.

\begin{flushleft}
(b) Case $1/2 < k_L^* < 1$.
\end{flushleft}

Here the eigenvalues $\lambda_{\pm}$ form a complex conjugate pair.
Again it is convenient to introduce a new variable $\theta$ by
$\tan \theta = \sqrt{4 k^2 -1}$.  Using $\theta$, (\ref{GL})
can be expressed as
\beqar \label{GL2}
G_{0,L}(k) &=& \frac{ \sin \theta}{\sin [(L+1) \theta] - k \sin[L \theta]}
\eeqar
and the fixed point equation (\ref{fixed_point_general}) as
\beqar \label{fixed_point_trig}
\tan(L \theta_L^*) &=& \frac{[(k+1)-2 \cos \theta] \sin \theta}
{\cos 2 \theta - \cos \theta + k (1-\cos \theta)}
\;\; |_{k=k_L^*, \theta=\theta_L^*},
\eeqar
where $\theta_L^*$ is the value of $\theta$ at $k=k_L^*$.
Now in the limit $k^* \rightarrow (\frac{1}{2})^{+}$ we can use a small
$\theta^*$ expansion. Since the right-hand side of (\ref{fixed_point_trig})
diverges as $\theta^* \rightarrow 0$, small values of $\theta^*$ are
achieved only in the large $L$ limit (as expected). Thus, the aforesaid limit
is equivalent to the limit $L \rightarrow \infty$.

A $1/L$ expansion of (\ref{fixed_point_trig}) and use
of the relation $\tan^{-1}x=\pi/2-\tan^{-1}1/x$ yields, after some
algebra,
\begin{eqnarray} \label{angle_exp}
\theta^* & = & \frac{ \pi}{2 }\frac{1}{L} - \frac{5 \pi}{4}\frac{1}{L^2}
+ \frac{25 \pi}{8} \frac{1}{L^3}+ O( \frac{1} {L^4}) \;\;.
\end{eqnarray}
This translates, in terms of the fixed point value, to
\begin{eqnarray} \label{fixed_point_exp}
k_L^* & = & \frac{1}{2} + ( \frac{ \pi}{4})^2 \frac{1}{L^2} +
O( \frac{1}{L^3}) \;\;.
\end{eqnarray}

The critical value $k_c$ is obtained exactly from $k_L^*$ in the
$L \rightarrow \infty$ limit as expected.
However the exponent $\nu$ does {\it not} have the correct
limiting value (which is $1/2$) as we show now.

Linearizing the recursion relation (\ref{rel_G_G'_2})
around the fixed point, we get the {\it eigenvalue}
\beqar \label{aut_max}
\Lambda &=& \frac{\partial k'}{\partial k}|_{k=k^*(L,L')} =
\frac{\partial G_{0,L}(k)/\partial k}{\partial G_{0,L'}(k')/\partial k'}
|_{\stackrel{L'\leq L-1}{k=k'=k^*(L,L')}} \nonumber \\
&=&\frac{\partial G_{0,L}(\theta)^{-1}/\partial \theta}
{\partial G_{0,L'}(\theta')^{-1}/\partial \theta'}
|_{\stackrel{L'\leq L-1}{\theta=\theta'=\theta^*(L,L')}} \;\;.
\eeqar
Using the expression (\ref{GL2}) and the large $L$ expansions
(\ref{angle_exp}) and (\ref{fixed_point_exp}) for renormalization from
a cell of size $L+1$ to one of size $L$, we get after
some straightforward but lengthy algebra,
\beqar \label{der_G}
\frac{\partial G_{0,L}(\theta)^{-1}}
{\partial \theta}|_{\theta=\theta_L^*}
&=& -(\frac{2}{\pi^2}) L^2 [ 1 + (5- \frac{\pi^2}{8}) \frac{1}{L} +
O(\frac{1}{L^2})] \;\;.
\eeqar
This serves as the denominator in (\ref{aut_max}), while the numerator
must be calculated by substituting $L+1$ for $L$ in Eq.\ (\ref{GL2}),
differentiating as in (\ref{aut_max}), and then substituting
the expansions (\ref{angle_exp}) and (\ref{fixed_point_exp}).
Thus we finally obtain
\beqar \label{aut_max_2}
\Lambda_L &=& 1 + (2+ \frac{\pi^2}{4}) \frac{1}{L} + O(\frac{1}{L^2})
\eeqar
and
\beqar \label{nu_fin}
\nu &=& \lim_{ L \rightarrow \infty} \frac{ \ln [(L+1)/L]}
{ \ln \Lambda_L} = \frac{1}{2 + \pi^2/4} = 0.2238...
\eeqar

This exactly calculated exponent $\nu$ therefore badly overshoots
the correct value $\nu=1/2$ in the same fashion
as in the numerical evaluation of the previous section.
Up to this point, however, it could still be a consequence of
the finite size $L$ imposed by the boundary condition (\ref{bc2}).
In the next subsection we will remove one of the boundaries and work
with a semi-infinite one-dimensional system.

\subsection{Generating function solution for the semi-infinite system}
\label{sec:semi-infinite}

Let us again consider a one-dimensional lattice where $x=0,1,2,....$ and
the site $x=-1$ has an infinitely repulsive barrier (see Fig.\ \ref{fig3}),
but there is no longer a barrier at the other end.
For this calculation we will exploit a different method, namely
the Laplace-Fourier method (see, e.g. second reference in \cite{HK}).

The master equation for the bulk is:
\beqar \label{master_bulk}
P_{0,x}(N+1) &=& \frac{1}{2}[P_{0,x-1}(N)+P_{0,x+1}(N)]
\eeqar
for $ x =1,2,...$ and the boundary condition at the {\it surface} $x=0$ is:
\beqar \label{bc3}
P_{0,0}(N+1) &=& \frac{1}{2} P_{0,1}(N)
\eeqar
We introduce the Fourier transform for the semi-infinite line by
\beqar \label{fourier_1d}
\Ptilde(q,N) &=& \sum_{x=0}^{+ \infty} e^{-iqx} P_{0,x}(N)
\longleftrightarrow P_{0,x}(N) = \intpi e^{iqx} \Ptilde(q,N) \;\;.
\eeqar
Multiplying (\ref{master_bulk}) by $e^{-iqx}$ and summing from $x=1$
to $\infty$, and using the boundary condition (\ref{bc3}), we get
\beqar \label{momentum_space}
\Ptilde(q,N+1) &=& \Ptilde(q,N)\cos q - \frac{1}{2}e^{iq}P_{0,0}(N) \;\;.
\eeqar
We now Laplace transform in $N$ by defining the generating function,
\beqar \label{laplace_1d}
\Gtilde(q,\lambda) &=& \sum_{N=0}^{\infty} \lambda^N \Ptilde(q,N)
=\sum_{x=0}^{\infty} e^{-iqx} G_{0,x}(\lambda)
\eeqar
where $G_{0,x}(\lambda)$ is the generating function for the
probability $P_{0,x}(N)$. Then from equation (\ref{momentum_space})
and the initial condition $P_{0,x}(0)= \delta_{0,x}$ we get
\beqar \label{gtilde_1d}
\Gtilde(q,\lambda) &=& \frac{1- \frac{1}{2} \;
\lambda e^{iq} G_{0,0}(\lambda)} {1-\lambda \cos q} \;\;.
\eeqar
This allows the determination of $G_{0,0}(\lambda)$. Integrating
(\ref{gtilde_1d}) over the first Brillouin zone, we get
\beqar \label{G0_1d}
G_{0,0}(\lambda) &=& \frac{I_0(\lambda)}
{1+ \frac{1}{2} \lambda I_1(\lambda)}
\eeqar
where we have defined the integrals
\beqar \label{Ix_def}
I_{x}(\lambda) &=& \intpi \frac{e^{iqx}}{1- \lambda \cos q}
\eeqar
for $x=0,1,2,..$. This integral can be easily computed as
a contour integral
\beqar \label{Ix_contour}
I_{x}(\lambda)&=& \frac{i}{ \pi } \oint_{C:|z|=1} dz \;
\frac{z^x}{\lambda (z - z_{+})(z-z_{-})}
\eeqar
where
\beqar \label{poles}
z_{\pm} &=& \frac {1 \pm \sqrt{1- \lambda^2}}{\lambda} \;\;.
\eeqar

It is worth mentioning that these poles are the same as the
eigenvalues $\lambda_{\pm}$ derived in the previous subsection,
since the relation between the two fugacities is $\lambda=2 k$.
Again the case $0<\lambda <1$ and $1 <\lambda <\infty$ have to
be distinguished. In the first case the poles $z_{\pm}$
lie on the real axis and $z_{-}$ is always interior to
the unit circle, while $z_{+}$ is always exterior.
In the second case the poles $z_{\pm}$ are complex conjugate
and lie on the unit circle symmetrically with respect to
the real axis. The result for the integral is
\beqar \label{Ix_result}
I_{x}(\lambda) &=& \left\{ \begin{array}{ll}
                I_0(\lambda) z_{-}^x \; \; & \mbox{if $0<\lambda <1$} \\
                -\frac{1}{\lambda} \frac{z_{+}^x-z_{-}^x}{z_{+}-z_{-}}
\; \; & \mbox{if $ 1< \lambda < \infty$}
                \end{array}
                \right.
\eeqar
where
\beqar \label{I0}
I_{0}(\lambda) &=& \left\{ \begin{array}{ll}
                1/\sqrt{1-\lambda^2}\;\; & \mbox{if $0<\lambda <1$} \\
               0 \;\; & \mbox{if $ 1< \lambda < \infty$}
                \end{array}
                \right.
\eeqar

We now proceed to compute the $G_{0,x}(\lambda )$ in the two cases. From
Eq.\ (\ref{gtilde_1d}) and (\ref{G0_1d}), we find
by inverse transforming in $q$,
\beqar \label{Gx_1d_real}
G_{0,x}(\lambda) &=& \frac{I_x(\lambda)}{1+\frac{1}{2} \lambda z_{-}
I_0(\lambda)}
\eeqar
in the case $0< \lambda <1$, and
\beqar \label{Gx_1d_complex}
G_{0,x}(\lambda) &=& I_x(\lambda)
\eeqar
for the case $1< \lambda < \infty$.

Consider first the case of $1 < \lambda <\infty$.  In this case, we
introduce an angle $\theta$ by
\beqar \label{angle_theta}
z_{\pm} &=& e^{\pm i\theta} \;\;,\; \;
\tan \theta = \sqrt{\lambda^2 -1} \;\;.
\eeqar
Then, from (\ref{Ix_result}) and (\ref{Gx_1d_complex}), we get
\beqar \label{G-1_complex}
G_{0,L}(\lambda)^{-1} &=& - \frac{\sin \theta}
{\cos\theta \sin L\theta} \;\;.
\eeqar
For large $L$, this expression changes sign rapidly as $\theta$
(or $\lambda$) varies, and thus unacceptable as the correlation
function on physical grounds.

Therefore, we now consider the case $0<\lambda < 1$.
If we define, as before, an angle $\alpha$ from
\beqar \label{angle_alpha}
z_{\pm} &=& e^{\pm \alpha} \;\;,\;\;
\tanh \alpha = \sqrt{1-\lambda^2} \;\;,
\eeqar
we find
\beqar \label{G-1}
G_{0,L}(\lambda)^{-1} &=& e^{\alpha L} [ \tanh \alpha +
\frac{e^{-\alpha}}{2 \cosh \alpha} ] \;\;.
\eeqar
We note that this is an exact result independent of any
renormalization procedure.  Now since the only singularity of
$G_{0,L}(\lambda )$ is at $\lambda =1$ (or $\alpha =0$) where
$\frac{\partial G_{0,L}}{\partial\lambda}$ diverges, we must interpret
$\lambda =1$ to be the critical point $\lambda_c$.  Thus,
it is also consistent with the mean field theory for semi-infinite
systems where the transverse correlation at criticality behaves
as the separation $L$ goes to $\infty$ as $L^{-(d-1)}$ for $d$
dimensions.

Now, applying the idea of cell renormalization discussed before to
this result, we turn to the equation for the fixed point,
\beqar \label{fixed_1d}
G_{0,L'}(\lambda^*) &=& G_{0,L}(\lambda^*)
\eeqar
where $L' \leq L-1$ and $L=1,2,...$.
This gives the fixed point at $\alpha^*=0$ or $\lambda^*=1$ exactly
independent of $L$ or $L'$, which is consistent with $\lambda_c =1$.
Turning to the critical exponent, we calculate the {\it eigenvalue}
\beqar \label{Lambda}
\Lambda &=& \frac{\partial \lambda'}{\partial \lambda}|_{\lambda^*}
= \left. \frac{\frac{\partial G_{0,L}(\lambda)}{\partial \lambda}}
{\frac{\partial G_{0,L'}(\lambda')}
{\partial \lambda'}} \right|_{\lambda^*}
\eeqar
by linearizing the recursion relation
around the fixed point $\lambda^*=1$. Note that, since
$G_{0,L}(\lambda)$ is singular at $\lambda=1$, what we are attempting
to do is to linearize around a singular fixed point.
After some manipulation we find
\beqar \label{Lambda_reusult}
\Lambda (L,L') &=& \frac{L+1}{L'+1} \;\;,
\eeqar
which is clearly wrong, since it would yield $\nu=1 + 0(1/L)$!

Even though the renormalization approach based on Eq.\ (\ref{rel_chi})
with (\ref{G-1}) results in a nonsensical exponent value,
the result for the two-point correlation function
$G_{0,L}(\lambda)$ itself is correct. Indeed it is easy to check that
it reproduces the exact results for the critical exponents
$\nu$ and $\gamma_1$ (a surface exponent \cite{Binder})
if they are calculated directly from it.
More specifically one finds in the grand-canonical ensemble,
\beqar \label{susc1_1d}
\chi_1(\lambda) &=& \sum_{L=0}^{\infty} G_{0,L}(\lambda)
\stackrel{\lambda \rightarrow 1^{-}}{\sim} (1-\lambda)^{-1/2}
\eeqar
which is the exact result for the surface exponent $\gamma_1=1/2$
(see next section) and
\beqar \label{r2_1d}
\xi^2(\lambda) &=& \frac{\sum_{L=0}^{\infty} L^2 G_{0,L}(\lambda)}
{\sum_{L=0}^{\infty} G_{0,L}(\lambda)}
\stackrel{\lambda \rightarrow 1^{-1}} {\sim} (1-\lambda)^{-1}
\eeqar
which again gives the exact result $\nu=1/2$.  The latter result can
also be read off directly from (\ref{G-1}) by noting that
$\alpha \sim \sqrt{1-\lambda }$ near $\lambda^*=1$.

So what is wrong with applying Eq.\ (\ref{rel_chi}) to our
$G_{0,L}(\lambda )$? In the present calculation, $L$ is
not the size of the system (which is infinite), but it refers only
to a site of the semi-infinite, one-dimensional lattice.
Therefore, the ideas of cell renormalization has no basis of
application in this case.  In fact, for general dimension $d$,
the transverse correlation at criticality should behave as
\beqar \label{correlation_function_1d}
G_{0,L}(\lambda)|_{\lambda=\lambda_c} &\sim&
\frac{1}{L^{d-2+\eta_{\perp}}}
\eeqar
for $L$ large. Since $\eta_{\perp}=1$ for random walks, such behaviour
cannot be consistent with Eq.\ (\ref{rel_chi}) for any $d$ other than
$1$.  (We just saw that it does not work even for $d=1$.)

If we assume instead
\beqar \label{rel_G_G'_new}
G_{0,L'}(\lambda')&=& (\frac{L'}{L})^a G_{0,L}(\lambda)
\eeqar
for an unknown exponent $a$, then this would lead to
\beqar \label{fixed_exact}
\lambda^* &=& 1 - O(\frac{1}{L}) \;\;, \\
\nu &=& \frac{1}{a+1} + O(\frac{1}{L}) \;\;,
\eeqar
for renormalization from $L+1$ to $L$.
The choice of $a=1$ then leads to the correct limiting values
as $L \rightarrow \infty$.  (This is also true for any ratio
$L/L'$ as shown in the Appendix.) The extra factor $(L/L)^a$
could be considered to correspond to the {\it rescaling} step
of the renormalization transformation.  However, this choice of $a$
is not in agreement with the intuitive guess of $-d+2-\eta_{\perp}$
for $d=1$.  Rather, it would correspond to $-d+2-\eta$ where $\eta=0$
is the bulk exponent.  Thus it is not straightforward to repair this
approach in a satisfactory way, even where {\it exact} correlation
functions are available, not to mention that in most cases we do
not have such luxury.

\section{Finite size scaling and surface critical behaviour}
\label{sec:finite_size}
We learned from the previous calculations that, in the usual form,
the cell renormalization procedure cannot be consistent. That is,
it is not assured that, as the size of the cell increases, the results
for the critical exponents improve and become exact in the limit
of an infinite cell.  If an improvement is attempted by using
exact correlation functions, an inconsistency is again found
stemming from the basic recursion relation (\ref{rel_chi}) itself.
On the other hand one might still hope that, as the cell size grows,
the {\em approximate} renormalization procedure of FG's truncation
recipe could give better and better results.  Unfortunately, however,
this also appears not to be the case.  Therefore, we need an
alternative procedure which is both consistent in principle and
workable in practice.

The clue of where to start comes from the result (\ref{fixed_point_exp})
where we calculated {\em how} the fixed point was becoming exact
in the $L \rightarrow \infty$ limit.
This is indeed compatible with the finite size scaling hypothesis
(see e.g \cite{Binder} and references therein),
\beqar \label{kc(L)_1}
k_c(L) &=& k_c(\infty) + A(\frac{1}{L})^{1/\nu}
\eeqar
where $k_c(\infty)=1/2$ is the exact critical fugacity in the
infinite lattice limit and $\nu=1/2$ in this case.
The idea, therefore, is that one can estimate $k_c(L)$ by looking
at the divergences of the (bulk) susceptibility defined as
\beqar \label{chi_bulk}
\chi_B(k) &=& \frac{1}{|\Lambda|} \sum_{\0 \in \Lambda}
\sum_{N=0}^{\infty} C(\0,N; \Lambda) k^N  \nonumber \\
         &=&  \frac{1}{|\Lambda|} \sum_{\0 \in \Lambda}
\sum_{\x \in \Lambda} G_{\0,\x}(k)
\eeqar
where $C(\0,N;\Lambda )$ is the number of $N$-step walks
starting from a point $\0$ and entirely contained in the volume
$\Lambda$ and $G_{\0,\x}(k)$ is its generating function.
The subscript $B$ refers to {\it bulk} in the sense that
the endpoints $\0$ and $\x$ can be anywhere in volume $\Lambda$.
Then, either by fixing the exact value of $\nu=1/2$ one can calculate
$k_c(\infty)$ or by fixing the exact value of $k_c(\infty)$
one can calculate the value of the exponent $\nu$.

The results for the square lattice
are shown in Fig.\ref{fig4} and Fig.\ref{fig5},
and they are consistent with the expected values. Indeed a best fit
for both cases gives  $k_c(\infty)=0.25 \pm 0.01$ and
$1/\nu=1.94 \pm 0.02$ and in the case of $\nu$ improves if we
include more and more terms corresponding to larger cell sizes.

The presence of the surface also changes the entropic
critical exponents as it is well known \cite{Binder}.
Indeed if the system is sufficiently large to make the distinction
between surface and bulk sensible, one can decompose the total
free energy $F_L$ as:
\beqar \label{free_total}
F_L(\Delta k,h,h_1,\frac{1}{L}) &\stackrel{L>>1}{\approx}& L^{d}
                     f_B(\Delta k,h,\frac{1}{L})
                   + L^{d-1} f_S(\Delta k,h,h_1,\frac{1}{L})
\eeqar
where $h$ and $h_1$ are the external fields associated with the bulk
and the surface respectively and $\Delta k=k-k_c$.
By differentiating twice with respect to the proper external field
and by using the finite size scaling {\em ansatz}, one gets the
well known general results
\beqar \label{chi_scaling}
\hat{\chi}_B(L) &\stackrel{L>>1}{\sim}& L^{\gamma /\nu} \nonumber \\
\hat{\chi}_1(L) &\stackrel{L>>1}{\sim}& L^{\gamma_1 \nu} \\
\hat{\chi}_{1,1}(L) &\stackrel{L>>1}{\sim}& L^{\gamma_{1,1} \nu} \;\;,
\nonumber
\eeqar
at the critical values $\Delta k=0$, $h=h'=0$.
Here we have defined the {\em local} susceptibilities:
\beqar \label{surf_susceptibility_rw}
\chi_1(k) &=& \frac{1}{|\partial \Lambda|}
\sum_{\0 \in \partial \Lambda}
\sum_{\x \in \Lambda} G_{\0,\x}(k)  \nonumber \\
\chi_{1,1}(k) &=& \frac{1}{|\partial \Lambda|}
\sum_{\0 \in \partial \Lambda}
\sum_{\x \in \partial \Lambda} G_{\0,\x}(k)
\eeqar
where we mean by $\partial \Lambda$ the boundaries of the volume
$\Lambda$.

Thus, our new method forgoes the usual corner rule {\it renormalization}
per se, and instead, calculates various quantities associated with a
finite cell and interprets them in terms of {\it surface} finite size
scaling.  From Fig.\ref{fig6} one can see that this method very
accurately reproduces the values predicted by the mean field
theory \cite{Binder,MST}, namely
$\gamma/\nu=2$, $\gamma_1/\nu=1$, $\gamma_{1,1}/\nu=-1$.

\section{Phase diagram for surface-bulk random walks}
\label{sec:bulk-surface}
As an application of the method just described, we now
present the finite size scaling solution of the problem
of the interplay between the bulk $\Lambda$ (with fugacity $k$)
and the surface $\partial \Lambda$ (with fugacity $k_1$)
(see Fig.\ref{fig7}),
based on the exact calculation of the finite cell susceptibilities.

Physically the possibility of changing the strength of the surface
fugacity with respect to the strength of the bulk fugacity
allows the surface to {\it make up} for the missing bonds.
Clearly one expects that if $k_1$ is sufficiently strong almost
all walks lie on the surface, and then the critical point and
the universality class should both change: when all the "interactions"
in the bulk are zero, the walks are not allowed to stay in the bulk
and we have the {\em adsorbed } phase.
Since $k_1=1/2$ and $k=1/4$ are the exact critical values corresponding
to an infinite {\it surface} (a line in this case) and an infinite bulk,
one then expects a qualitative phase diagram as shown in Fig.\ref{fig8}.

For this calculation we eliminated the {\it corner} and imposed
a periodic boundary condition into the vertical direction $\hat{2}$,
while the horizontal direction $\hat{1}$ is of size $L$ and
has free edges.  The result is shown in Fig.\ref{fig9}.
It appears that there is a tricritical point (called
{\em special point}) which is the intersection of three different
lines (corresponding to three different second order phase transitions).
Below the special point there is the {\em ordinary} transition,
where the bulk and the surface undergo a transition at the
same critical point. Above the special point there is a line of
{\em surface} transitions, which take place if $k_1$ is bigger then the
special point ordinate, where the surface goes into an {\it ordered}
state (where the susceptibility is singular) while
the bulk is still {\it disordered}, as well as another line
called the {\em extraordinary} line where the bulk also becomes
{\it ordered}.

Our estimate of the special point $SP$ is at $k=0.25 \pm 0.01$,
and $k_1=0.35 \pm 0.01$, corresponding to a ratio $k_1/k=1.40 \pm 0.06$.
This is in reasonably good agreement with a simple mean-field argument
which would predict the ratio of $4/3$. The errors were estimated
graphically.

The qualitative features contained in this phase diagram also
appear in the case of the self-avoiding walks \cite{nakanishi}
and in percolation \cite{de'bell and essam}.

Quite similar features are found in the case of a defect (which is
a $d-1$-dimensional surface inserted into a $d$-dimensional
bulk). The phase diagram obtained looks very similar to the one
for the surface. The special point $SP$ is found when $k_1/k=1$
again consistent with mean field arguments.
\section{Conclusions}
\label{sec:conclusions}
In this paper we have given a detailed analysis of the difficulties
associated with the usual cell renormalization approach
to the random walk problem and presented an alternative method
to calculate critical properties which does not suffer from similar
difficulties.  This new approach is shown to give results which improve
as the size of the cell increases, unlike the previous approaches
which are shown here not to have this essential feature,
relying on an uncontrolled approximation.
Our approach is based only on the finite size scaling hypothesis.
Using this approach we have computed the full phase diagram of the
effect of the surface fugacity having a different value from the bulk
fugacity, and calculated the exponents $\gamma_{1}$ and $\gamma_{1,1}$
as well $\nu$.  All the results are consistent with simple mean field
arguments as expected. Analogous results are obtained for the
case of a defect.

The real challenge now is to use this method for the situations where
no simple mean field results are useful and the exponents are unknown.
A timely example of such a case is the one where the substrate
on which the random walks are constrained is disordered or otherwise
self-similar.  Although this problem is potentially important as it
has essential features of transport through disordered media, an
important materials problem, the results of previous investigations
\cite{GK,SJ} are not reliable. This extension is not expected to be
straightforward, however, since
various numerical approaches \cite{GNMF,GM,GMN,GNF} all agree on the
result that the susceptibility singularity for an {\em ideal chain}
or {\em trapped ant} (which can also be thought as the limit
of the self-avoiding walk in absence of self-avoidance) is not
a simple power law in the presence of strongly correlated disorder,
but rather an {\em essential singularity}.
Further work will then be needed in order to implement this extension.

\acknowledgments
We are deeply indebited with Amos Maritan who suggested us
the problem and in particular the calculation of Sec.III A.
Enlighting discussions with Attilio Stella and Flavio Seno
are gratefully acknowledged. One of us (A.G.) wishes to thank the
Department of Physics at the University of Padova for the kind
hospitality and INFN sez di Padova during several visits in the
course of this work and also for partial financial support.

\newpage
\appendix
\section*{Exact decimation for an arbitrary scale factor }
We will show in this Appendix that, unlike in the {\em corner rule}
renormalization, an exact decimation procedure (which
gives exact scaling relations) always gives the exact critical point
and the exact critical exponent irrespective of the choice
of the scaling ratio $l \equiv L/L'$.
Without loss of generality, we will do this for decimation from
$L$ to $a=1$ ($a$ being the lattice constant), with the rescaling ratio of $L$.

It is not hard to convince oneself that a decimation of the $L$ sites
next to the origin gives rise to the recursion,
\beqar \label{rec_k_ex}
k'_L &=& \frac{k^L}{1-2k^2 A_L(k)} \prod_{p=1}^L A_p(k) \;\;,\;\;\;
         (k \equiv k_1 )
\eeqar
and
\beqar \label{rec_G_ex}
G_{0,x'=x/L}(k') &=& (1-2k^2 A_L(k)) G_{0,x}(k)
\eeqar
where $A_L(k)$ is such that
\beqar \label{rec_A_ex}
A_{L+1} &=& \frac{1}{1-k^2 A_L} \;\;.
\eeqar
In the previous expressions $k_L$ refers to the fugacity $k$
when the lattice constant is $L$.
We can show by induction that $k^*_L = \frac{1}{2}$ ($\forall L \in$
{\fancy N}). Indeed if we assume
$k^*_L = k^*_1 = \frac{1}{2} \equiv k^*$, then we would have
from Eq.\ (\ref{rec_k_ex}),
\beqar
k^*_{L+1} &=& k^*_L D_L(k^*)
\eeqar
where
\beqar \label{fixed_point_D}
D_L(k^*) &=& [k A_{L+1}(k) \frac{1-2k^2 A_L(k)}
             {1-2k^2 A_{L+1}(k)}]_{k=k^*} \;\;.
\eeqar
Thus the claim is equivalent to showing that $D_L(k^*) = 1$,
which is easy to derive using Eq.\ (\ref{rec_A_ex}).

Next we will show, again by induction, that $\nu = \frac{1}{2}$
independent of $L$.
{}From
\beqar \label{recursions_k}
k'_{L+1} &=& k'_{L} D_L(k) \;\;.
\eeqar
Upon differentiation with respect to $k$, we have from Eq.\
(\ref{recursions_k}),
\beqar \label{derivate_k}
\frac{\partial k'_{L+1}}{\partial k}
        &=& \frac{\partial k'_{L+1}}{\partial k} D_L(k)
         +  k'_{L} \frac{\partial D_L(k)}{\partial k}
\eeqar
Since by hypothesis,
\beqar
\frac{\partial k^{\prime}_L}{\partial k}|_*
     &=& \lambda_{MAX}(L) [D_L(k)]_{k=k^*}
      +  k^*_L [\frac{\partial D_L(k)}{\partial k}]_{k=k^*} \;\;,
\eeqar
after some manipulations we get, at the fixed point $k^* = \frac{1}{2}$,
\beqar
D_L(k^*) &=& 1 \;, \; \frac{\partial D_L(k)}{\partial k} |_*
    = 2(2L+1) \;\;.
\eeqar
Therefore, substituting in (\ref{derivate_k}), we get
\beqar
\frac{\partial k'_{L+1}}{\partial k} |_* &=& (L+1)^2
      = \lambda_{MAX} (L+1)
\eeqar
which is what we wanted to prove.
%
%

%
%
\begin{figure}
\caption{Corner rule for the case of the smallest cell $2a \times 2a$.
By symmetry the number of walks starting from the origin $0$ and
exiting in the direction $\hat{1}$ or $\hat{2}$ are equal.}
\label{fig1}
\end{figure}
\begin{figure}
\caption{Comparison between the exact and approximate
$G_L^{\uparrow}(k) \equiv \chi_{\hat{2}}(k,1/L)$ in the cases $L=1,2,3$.
The three solid lines correspond to the exact
evaluations, while the two dotted lines are the approximate results
as discussed in text. The intersections
of the the solid lines correspond to the exact fixed points
$k^*=0.3157$, $0.2950$, while the intersections of the dotted lines
correspond to the Family-Gould fixed points $k_{FG}^*=0.3470$, $0.3108$
for a scaling from $L=2,3$ to $L'=1,2$ respectively.}
\label{fig2}
\end{figure}
\begin{figure}
\caption{One-dimensional case. In (a) the sites $x=-1$
and $x=L+1$ have an infinitely repulsive barrier, corresponding
to the corner rule in $d=1$. The second case (b) is a semi-infinite
one-dimensional lattice with an infinitely repulsive barrier only
at $x=-1$.}
\label{fig3}
\end{figure}
\begin{figure}
\caption{Finite-size scaling result for the critical value $k_c(\infty)$,
which corresponds to the true critical point for the square lattice.
A best fit over all points gives $k_c(\infty)=0.25 \pm 0.01$, while
the exact value is $k_c=1/4$.}
\label{fig4}
\end{figure}
\begin{figure}
\caption{Finite-size scaling result for the exponent $\nu$.
A best fit over all points gives $1/\nu=1.94 \pm 0.02$, while
the exact value is $1/\nu=2$.}
\label{fig5}
\end{figure}
\begin{figure}
\caption{Evaluation of the bulk and surface susceptibilities
in the finite-size scaling approach. The estimates for
$\chi_B$ ($\bigcirc$), $\chi_1$ ($\bigtriangleup$)
and $\chi_{1,1}$ ($+$) are obtained from the slopes of
the lines shown in the log-log plots.
The exact values according to the mean field calculation
are $\gamma/\nu=2$, $\gamma_{1}/\nu=1$ and $\gamma_{1,1}/\nu=-1$.}
\label{fig6}
\end{figure}
\begin{figure}
\caption{Example of a surface-bulk problem where the surface fugacity
$k_1$ and the bulk fugacity $k$ are shown.}
\label{fig7}
\end{figure}
\begin{figure}
\caption{A sketch of the expected phase diagram.
The lines shown correspond to the ordinary transition (O), the surface
transition (S) and the extraordinary transition (E). Also
shown is the tricritical point called the special transition (SP).}
\label{fig8}
\end{figure}
\begin{figure}
\caption{Computed phase diagram for the surface-bulk problem
of Fig.\protect{\ref{fig7}}.  The points shown have been calculated for
different
system sizes, $L=4$ ($\bigcirc$), $L=10$ ($\bigtriangleup$), and
$L=20$ ($\Diamond$). The dotted and dashed lines correspond to slopes
$k_1/k=1,2$ respectively. The special point is estimated to be at
$k=0.25 \pm 0.01$, $k_1=0.35 \pm 0.01$.}
\label{fig9}
\end{figure}
%
%
%

\begin{table}
\caption{Behaviour of $\nu$ as functions of the cell size
$b=L/a$ where $a$
is the lattice constant. The first two columns
refer to the present work, while the second two refer to the results
using the approximate recipe of Family and Gould ref\protect{\cite{FG}}.}
\begin{tabular}{lcccc}
\multicolumn{1}{l}{Scaling length $b/b'$}&
\multicolumn{1}{c}{$k^*$}&
\multicolumn{1}{c}{$\nu$}&
\multicolumn{1}{c}{$k_{FG}^*$}&
\multicolumn{1}{c}{$\nu_{FG}$} \\
\hline
$2/1$ &$0.3156$ &$0.5438$&$0.3470$ &$0.5853$   \\
$3/1$ &$0.2950$ &$0.5132$&$0.3108$ &$0.5571$   \\
$3/2$ &$0.2770$ &$0.4441$&$0.2920$ &$0.5129$   \\
$4/1$ &$0.2825$ &$0.4937$&$0.2926$ &$0.5412$   \\
$4/3$ &$0.2711$ &$0.4485$&$0.2743$ &$0.4868$   \\
$5/1$ &$0.2745$ &$0.4792$&$0.2838$ &$0.5398$   \\
$5/4$ &$0.2640$ &$0.4351$&$0.2693$ &$0.5148$   \\
\end{tabular}
\label{table1}
\end{table}

\newpage

\begin{references}
\bibitem{Wilson} K. Wilson and J. Kogut, Phys. Rep. {\bf 12 C}, 75 (1972)
\bibitem{Creswick} R. Creswick, H. Farach and C.Poole Jr.,
{\em Introduction to Renormalization Group methods in physics}
(Wiley, 1992)
\bibitem{Zinn-Justin} J. Zinn-Justin, {\em Quantum field theory
and critical phenomena} (Oxford, 1989)
\bibitem{Burkhardt} See, e.g., {\em Real-Space Renormalization}, ed.
T. Burkhardt and J.M.J. van Leeuwen (Springer, New York, 1982)
and references therein
\bibitem{FG} F. Family and H. Gould, J. Chem. Phys {\bf 80}, 3892 (1984)
\bibitem{HN} H. Nakanishi, J. Phys. A {\bf 17}, L329 (1984).
\bibitem{SJ} M. Sahimi and G.R. Jerauld, J. Phys. C  {\bf 16}, L1043 (1983)
\bibitem{GK} H. Gould and R.P. Kohin, J. Phys. A {\bf 17}, L159, (1984)
\bibitem{HK} See, e.g., J.W. Haus and K.W. Kehr, Phys. Rep. {\bf 150},
263 (1987); S. Havlin and D. Ben-Abraham, Adv. Phys. {\bf 36}, 695 (1987);
and references therein
\bibitem{deG} P.G. de Gennes, Recherche {\bf 7}, 919 (1976);
C. Mitescu and J. Roussenq, in { \it Percolation Structures
and Processes}, ed. G. Deutscher, R.Zallen and J.Adler,
Annals of the Israel Physical Society No.5 (Higler, Bristol,
1983), pg 81.
\bibitem{Maritan} A. Maritan, Phys. Rev. Lett. {\bf 62}, 2845 (1989)
\bibitem{GNMF} A. Giacometti, H. Nakanishi, A. Maritan and N.H. Fuchs,
J. Phys. A {\bf 25}, L461 (1992)
\bibitem{GM} A. Giacometti and A. Maritan, to appear in Phys.Rev. {\bf E}
(1993)
\bibitem{GMN} A. Giacometti, A. Maritan and H. Nakanishi, preprint
PURD-TH-93-05 (1993)
\bibitem{GNF} N. Fuchs, A. Giacometti and H. Nakanishi, preprint
PURD-TH-93-10 (1993).
\bibitem{Binder} K. Binder in {\em Phase Transitions and Critical
Phenomena} ed. C. Domb and J. Lebowitz, vol.8 (Academic Press,1983)
\bibitem{MW-65} E. Montroll and G. Weiss, J. Math. Phys {\bf 6}, 167 (1965)
\bibitem{Barber-Ninham} M. Barber and B. Ninham, {\em Random and
Restricted Walks: Theory and Applications} (Gordon and Breach, NY 1970)
\bibitem{ID} C. Itzykson and J.M. Drouffe {\it Statistical Field
Theory} vol.1 (Cambridge University Press, Cambridge 1989)
\bibitem{MST} These can also be computed directly; see, e.g.,
A. Maritan, A.L. Stella and F. Toigo, Phys.Rev B {\bf 40}, 9269 (1989)
\bibitem{nakanishi} H. Nakanishi, J. Phys. A {\bf 14}, L355 (1981)
\bibitem{de'bell and essam} K. de'Bell and J.W. Essam, J. Phys. C
{\bf 13}, 4811 (1980)
\end{references}
\end{document}